\documentclass[modern]{aastex61}

\usepackage{amsfonts}
\usepackage{amsmath}
\usepackage{amssymb}
\usepackage{upgreek}
\usepackage{bm}
\usepackage{dcolumn}
\usepackage{epsfig}
\usepackage{graphicx}
\usepackage{placeins}
\usepackage{comment}
\usepackage{gensymb}
\usepackage{lineno}
\usepackage{wrapfig}
\usepackage{ragged2e}
\usepackage[utf8]{inputenc}
\usepackage{enumitem}
\usepackage{color}
\definecolor{spring}{rgb}{0.7,0.9,0.7}
\definecolor{brick}{rgb}{0.7,0.2,0.1}
\definecolor{redHL}{rgb}{1.0,0.5,0.5}
\definecolor{darkgreen}{rgb}{0.0,0.5,0.0}

\usepackage{lipsum}

\def\cob#1{compact object binaries#1
  (COB#1)\gdef\cob{COB}}
\def\imbh#1{intermediate mass black hole#1
  (IMBH#1)\gdef\imbh{IMBH}}
\def\imbhb#1{intermediate mass black hole binary#1
  (IMBHB#1)\gdef\imbhb{IMBHB}}  
\def\bh#1{black hole#1
  (BH#1)\gdef\bh{BH}}
\def\bbh#1{binary black hole#1
  (BBH#1)\gdef\bbh{BBH}}
\def\gw#1{gravitational wave#1
  (GW#1)\gdef\gw{GW}}
\def\nr#1{numerical relativity#1
<argument> \mnras  (NR#1)\gdef\nr{NR}}
\def\snr#1{signal-to-noise-ratio#1
  (SNR#1)\gdef\snr{SNR}}
\def\pn#1{post-Newtonian#1
  (PN#1)\gdef\pn{PN}}
  \def\O1#1{observing run#1
  (O1#1)\gdef\pn{O1}}

\def \h70{{h_{70}}}
\begin{document}

\title{Astro2020 Decadal Science White Paper: What we can\\ Learn from Multi-band Observations of Black Hole Binaries}

\noindent
\begin{flushleft}
{\bf Main Thematic Area:} Formation and Evolution of Compact Objects\\
{\bf Secondary Thematic Area:} Multi-Messenger Astronomy and Astrophysics
\vspace{-1.8cm}
\end{flushleft}

\author{Curt Cutler} 
\affiliation{Theoretical Astrophysics, California Institute of Technology, Pasadena, CA 91125, USA}
\affiliation{Jet Propulsion Laboratory, 4800 Oak Grove Drive, Pasadena, CA 91109, USA}
\author{Emanuele Berti} 
\affiliation{Department of Physics and Astronomy, Johns Hopkins University, 3400 N. Charles Street, Baltimore, MD 21218, USA}
\author{Karan Jani}
\affiliation{Center for Relativistic Astrophysics and School of Physics, Georgia Institute of Technology, Atlanta, GA 30332, USA}
\author{Ely D. Kovetz} 
\affiliation{Department of  Physics and Astronomy, Johns Hopkins University, 3400 N. Charles Street, Baltimore, MD 21218, USA}
\author{Lisa Randall}
\affiliation{Department of Physics, Harvard University, 17 Oxford Street, Cambridge, MA 02138, USA}
\author{Salvatore Vitale}
\affiliation{LIGO, Massachusetts Institute of Technology, Cambridge, MA 02139, USA}
\affiliation{Department of Physics and Kavli Institute for Astrophysics and Space Research, MIT, Cambridge, MA 02139, USA}
\author{Kaze W.K. Wong} 
\affiliation{Department of Physics and Astronomy, Johns Hopkins University, 3400 N. Charles Street, Baltimore, MD 21218, USA}

\author{Kelly Holley-Bockelmann}
\affiliation{Department Physics and Astronomy, Vanderbilt University, 2301 Vanderbilt Place, Nashville, TN 37235, USA}
\affiliation{Department of Physics, Fisk University, 1000 17th Ave. N, Nashville, TN 37208, USA}
\author{Shane L. Larson}
\affiliation{CIERA, Northwestern U., Evanston, IL 60208, USA}
\author{Tyson Littenberg}
\affiliation{NASA Marshall Space Flight Center, Huntsville AL 35812, USA}
\author{Sean T. McWilliams}
\affiliation{Department of Physics and Astronomy, West Virginia University, Morgantown, WV 26506, USA}
\author{Guido Mueller}
\affiliation{University of Florida, Gainesville FL 32611, USA}
\author{Jeremy D. Schnittman}
\affiliation{Gravitational Astrophysics Laboratory, NASA Goddard Spaceflight Center, Greenbelt, MD 20771, USA}
\affiliation{Department of Physics, University of Maryland, Baltimore County, Baltimore, MD 21250, USA}
\author{David H. Shoemaker}
\affiliation{LIGO, Massachusetts Institute of Technology, Cambridge, MA 02139, USA}
\author{Michele Vallisneri} 
\affiliation{Theoretical Astrophysics, California Institute of Technology, Pasadena, CA 91125, USA}
\affiliation{Jet Propulsion Laboratory, 4800 Oak Grove Drive, Pasadena, CA 91109, USA}

\begin{abstract}
The LIGO/Virgo gravitational-wave (GW) interferometers have to-date detected ten merging black hole (BH) binaries, some with masses considerably larger than had been anticipated.  
Stellar-mass BH binaries at the high end of the observed mass range (with "chirp mass" ${\cal M} \gtrsim 25 M_{\odot}$) should be detectable by a space-based GW observatory years before those binaries become visible to ground-based GW detectors.  This white paper discusses some of the synergies that result when the same binaries are observed by instruments in space and on the ground.  We consider intermediate-mass black hole binaries (with total mass $M \sim 10^2 -10^4 M_{\odot}$) as well as stellar-mass black hole binaries.  We illustrate how combining space-based and ground-based data sets can break degeneracies and thereby improve our understanding of the binary's physical parameters.   
While early work focused on how space-based observatories can forecast precisely when some mergers will be observed on the ground, the reverse is also important: ground-based detections will allow us to "dig deeper" into archived, space-based data to confidently identify
black hole inspirals whose signal-to-noise ratios were originally sub-threshold, increasing the number of binaries observed in both bands by a factor of $\sim 4$--$7$. 
\end{abstract}

\thispagestyle{empty}

\setcounter{page}{1}
\section{Introduction} \label{sec:intro}

The LIGO/Virgo detections of gravitational waves (GWs) from (to-date)
ten binary black hole (BH) mergers and one binary neutron star merger
have already demonstrated the capacity of GW astronomy to yield
important and unexpected discoveries.

The first binary detected by LIGO, GW150914, would have been observable a few years prior to merger by a space-based detector, had one been flying then~\cite{2016PhRvL.116w1102S}.  Just as for multi-wavelength electromagnetic observations, combining information from different GW bands -- by observing the \emph{same source} as it sweeps through different frequency bands -- will enable novel ways to exploit the data.  

LIGO and Virgo are currently sensitive to GWs in the $\sim 30 - 1000\,$Hz band, and 
next-generation ground-based observatories will be sensitive down to  $\sim 1-10\,$Hz. Because lower frequencies imply longer detector arms, exploring the $\sim 10^{-4} - 10^{-1}\,$Hz regime requires an observatory in space.   Other Decadal white papers will describe diverse areas of astronomy that will be opened up by a low-frequency, space-based GW observatory: see e.g. \cite{BertiWP,ColpiWP,NatarajanWP,BellovaryWP,CaldwellWP,LittenbergWP,BakerWP,BerryWP,McWilliamsWP,CornishWP, EracleousWP}.
Here, we describe the science that will be enabled by GW observations of {\it the same} BH binaries from both ground and space.  

In addition to stellar-mass BHs of the sort detected so far, we will consider the implications for intermediate-mass BHs (IMBHs, $M \sim 10^2 - 10^4 M_{\odot})$. 
Electromagnetic observations have found some IMBH {\it candidates}~(see e.g. \cite{2009Natur.460...73F,2014Natur.513...74P}), but
the evidence for their existence is still not conclusive. 
{\bf GW astronomy promises the first cosmological survey of BHs in the full mass range from primordial to supermassive}, which should in turn help distinguish between various competing
BH formation scenarios in each different mass range. 

Section~\ref{sec:stellarbh} is mainly concerned with detection issues, while Sec.~\ref{sec:imbh} covers parameter estimation, and scientific inference more generally. 

\section{Multiband GW detection of BH Binaries} \label{sec:stellarbh}

Which binaries will be observable by both ground-based and space-based detectors?  To answer this question quantitatively, we need to choose some fiducial sensitivity curves.  For concreteness, we will use the planned Laser Interferometer Space Antenna (LISA) mission, with a noise power spectral density taken from \cite{Cornish:2018dyw} (but converted to standard noise conventions for ground-based detectors) as our fiducial example to illustrate what can be done from space. For ground-based detectors we will adopt
two proposed next-generation detectors: the Einstein Telescope (ET)~\cite{Punturo:2010zz} and the Cosmic Explorer (CE)~\cite{Evans:2016mbw}. 
For illustration, Fig.~1a depicts the signal strength of two fiducial binaries -- a 
$(30 + 30) M_{\odot}$ binary at $D = 100$~Mpc (i.e., a LIGO-type source) and a $(10^3 + 10^3) M_{\odot}$ IMBH binary at $z=1$ -- laid over the noise curves of LISA, a specific ET design called ET-D, and CE, respectively.  In this figure, signal strengths are plotted against signal frequency.
In the time domain, the inspiral signals are chirps with frequency increasing in time and, therefore, time increases  left-to-right along these tracks.

Both signal tracks begin four years prior to merger.  
We first see their inspiral in LISA, and roughly four years later
we see their merger and ringdown in ground detectors.
Fig.~1b shows the redshift $z$ and luminosity distance $D_L$ out to which BH binaries can be detected for these three observatories, as a function of their total mass. Here the binaries are taken to be equal-mass and non-spinning, the masses refer to the "source-frame," the detection threshold is taken as a signal-to-noise ratio (SNR) $ > 8$ for a source with optimal sky location and orientation, we use a canonical $\Lambda$CDM model to determine $D_L(z)$, and the gravitational waveforms 
are generated using the \texttt{IMRPhenomPv2} model~\citep{Hannam:2013oca,2015PhRvD..91b4043S}.

\begin{figure}[h]
\includegraphics[width=1\textwidth]{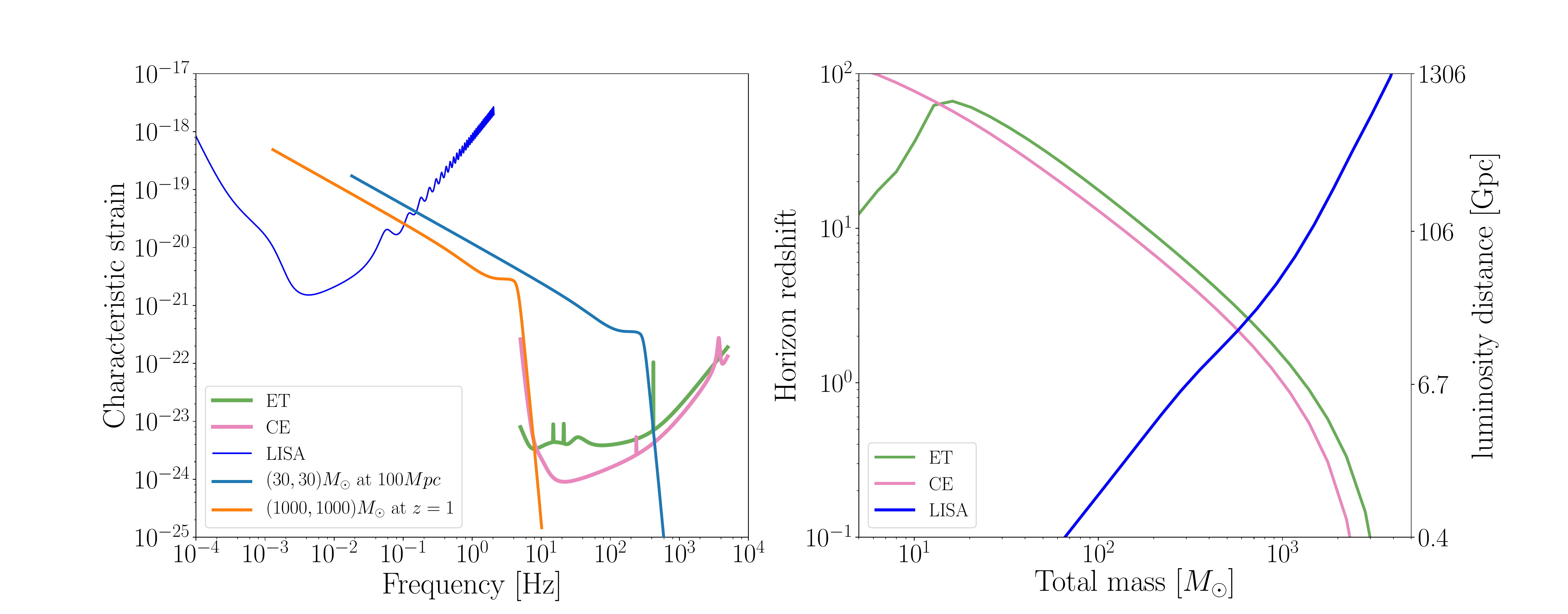}
\caption{{\it (a)}: Frequency tracks of binary BH signals, compared to the noise curves for LISA, ET and CE.
The $(30 + 30) M_\odot$ binary will radiate at $\sim10$ mHz four years prior to merger, while the $(1000 + 1000) M_\odot$ IMBH binary will sweep through most of the LISA band in that time. {\it (b)}: Redshift out to which BH binaries are detectable, for current and proposed ground-based GW experiments. We consider equal-mass, non-spinning BHs. Masses refer to the source frame, and our criterion for detectability is that the 
SNR should be $ > 8$ for a binary with optimal sky location and oriention.} \label{fig:1}
\end{figure}

Fig.~1b shows that for ET or CE, 
a GW150914-like signal will be detectable up to $z \gtrsim 10$.
The "sweet spot" for joint space and ground detection of IMBHs will be
$M \sim 700 M_{\odot}$; such binaries will be jointly detectable out to $z \sim 3$. 
For stellar-mass binaries, LISA will be much less sensitive than ET or CE, 
independently detecting only a small fraction of the stellar-mass binaries observed on the ground. 
However one can exploit the ground-based binary parameter measurements to greatly focus the search through archived LISA data, and hence reduce the SNR threshold needed to confidently identify the binary signals in the low-frequency data set. 
Wong et al.~\cite{2018arXiv180808247W} showed that this data analysis strategy should increase the number of multiband detections by a factor $\sim 4$--$7$. This increased number of detections will amplify the power of statistical comparisons between different astrophysical binary population and evolution models~\cite{Cholis:2016kqi,Gerosa:2017kvu,Stevenson:2017dlk,2018PhRvD..98h3017T,2019arXiv190200021G}.

\section{Enhanced Science from Multiband GW detections of BHB's}\label{sec:imbh}

Space-based GW observations will complement higher frequency Earth-based GW data in four main ways:

\noindent
{\bf 1) Space-based detections will provide hours to months of advance warning for some of the mergers that
will be observed on the ground.} 
This will alert GW observers to ensure their detectors should be "on" during the merger. While only a
small fraction of ground-based detections will benefit from such alerts, these will generally be the closest, loudest events, and so among the most useful ones for performing tests of general relativity, as discussed below. 
Naturally, this will also inform the EM community of an impending event. It is an open mystery whether these binary BH mergers would generate an EM counterpart. The general expectation is that the material left behind after the merger is not sufficient to power relativistic jets. However, the weak transient signal recorded by the Fermi GBM instrument less than a second after GW150914~\cite{Connaughton:2016umz} renewed interest in possible electromagnetic counterparts to binary mergers~\cite{BakerWP}. Space-based advance warnings will provide the merger time to within a minute, and an approximate sky location, thus helping electromagnetic astronomers prepare for the opportunity to open a new discovery space. 

\noindent
{\bf 2) Observations from space will give us access to features of the gravitational waveforms that may be absent (or poorly measured) in ground-based data.} A good example is orbital eccentricity. GW emission is well known to circularize binary orbits~\citep{Peters:1964zz}, and therefore  ground-based detectors are not expected to measure any significant eccentricity in most astrophysical scenarios.
A low-frequency, space-based interferometer will access an earlier phase of the inspiral, and it will measure the binary's eccentricity at a remarkable $\sim 0.01$ level~\citep{Nishizawa:2016jji}. This can be a powerful tool for distinguishing between BBH formation channels~\citep{2016ApJ...830L..18B,Nishizawa:2016eza,Rodriguez:2017pec,Randall:2018nud,DOrazio:2018jnv,Samsing:2018isx}.
Another example is the modulation of gravitational waveforms due to spin-induced precession of the orbital plane~\citep{PhysRevD.49.6274}, which similarly encodes important information on the binary formation mechanism~\cite{Gerosa:2013laa,Vitale:2015tea,Rodriguez:2016vmx,Farr:2017uvj,Farr:2017gtv,Gerosa:2018wbw}. Since ground-based detectors can only access the last few orbital precession cycles, they cannot usually measure individual spins with high precision~\citep{2013PhRvD..87b4035B,PhysRevLett.112.251101, PhysRevD.93.084042,Vitale:2016avz}. On the other hand, space-based missions  observe many more precession cycles. Preliminary work suggests that they might provide accurate spin measurements for IMBH binaries, which would in turn yield further insights into their formation mechanisms.

\noindent

\begin{wrapfigure}{r}{0.5\textwidth}
  \begin{center}
  \vskip -0.25in
    \includegraphics[width=0.5\textwidth]{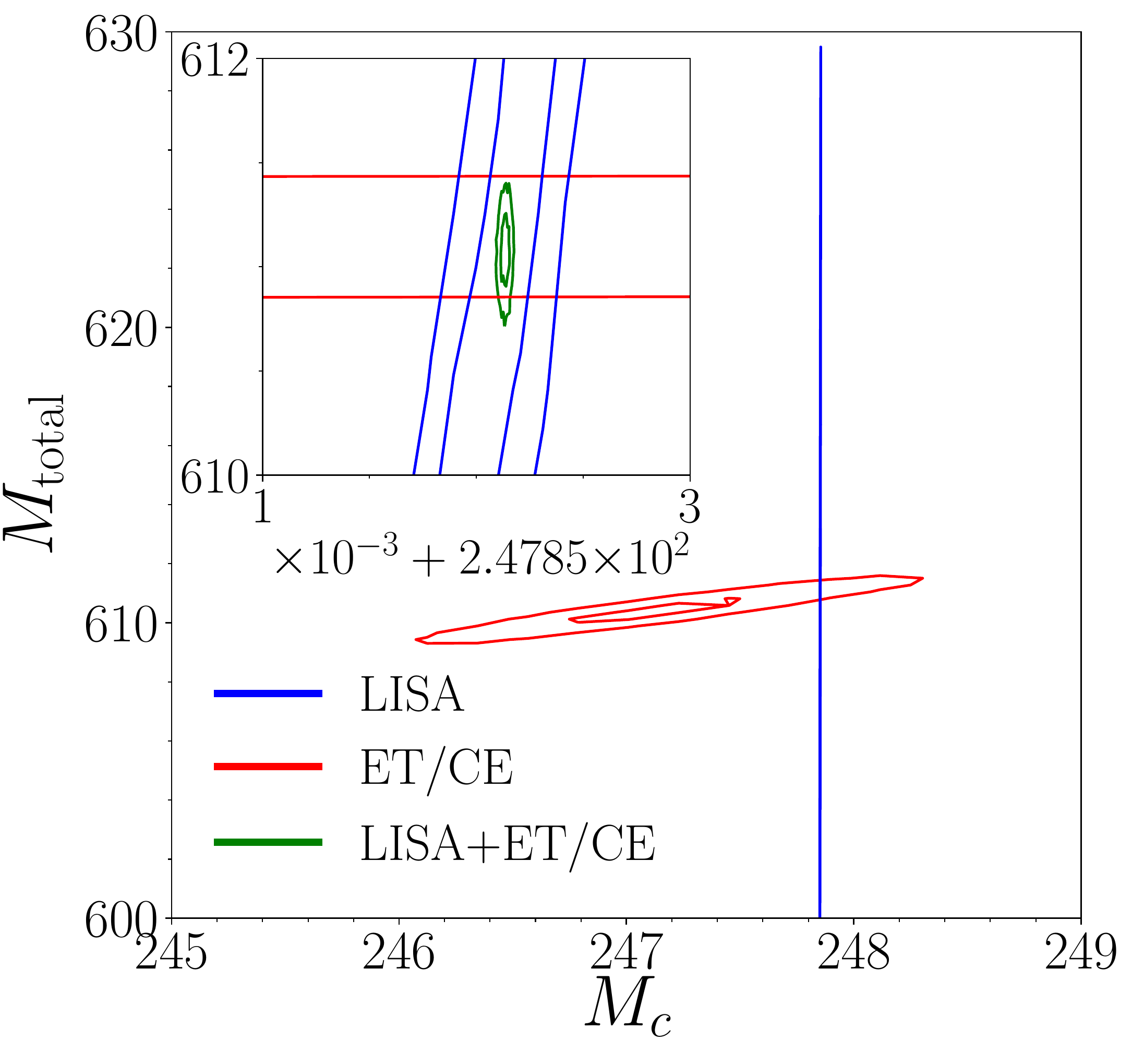}
  \end{center}
  \vskip -0.2in
  \caption{\small Posterior distribution of chirp mass $M_{c}$ and total mass $M_{\rm total}$ obtained using LISA (blue); a third-generation ground-based detector network consisting of ET~\citep{Punturo:2010zz} and CE~\citep{Evans:2016mbw} (red); and the combined measurement (dark green). The matched-filter SNR of this system was 5.5 in LISA and 1010 in the ground-based network. %
}
\label{fig:PE}
\end{wrapfigure}

{\bf 3) Multiband observations will yield complementary information that can remove degeneracies between parameters and/or improve parameter estimation.} 
This possibility was first suggested in Ref.~\cite{2016PhRvL.117e1102V} assuming aLIGO sensitivity.
In Fig.~\ref{fig:PE} we update that analysis by assuming -- more realistically, considering LISA's anticipated launch date  -- a ground-based detector network 
composed of one ET and one CE observatory.

Briefly, the results in Fig.~2 were derived as follows. 
We first ran a Fisher matrix calculation to produce posterior distributions representative of the LISA measurement.  In the Fisher matrix approximation, the likelihood is a multivariate Gaussian.
We augmented \textit{LALInference}~\citep{2015PhRvD..91d2003V}, the Bayesian stochastic sampler routinely used by the LIGO and Virgo collaborations, to use the covariance matrix produced in the first step as a Bayesian prior for parameter estimation with ground-based data.
We simulated the GW signals of IMBHs using the IMRPhenomPv2 waveform template~\citep{Hannam:2013oca,2015PhRvD..91b4043S} and added them to Gaussian noise. The final posterior was calculated using a coherent Bayesian analysis~\citep{2015PhRvD..91d2003V}. 
For this proof of principle, we neglected spins in our waveform models. Unlike in Fig.~1, the masses in Fig.~2 are "observer-frame" masses, and so larger than the "source-frame" values by the factor $(1+z)$. (This choice avoids the propagation of distance errors into the displayed result, which would unnecessarily complicate the interpretation.)  In this case, we see that the degeneracy-breaking shrinks the area of the 2-$\sigma$ error ellipse by about six orders of magnitude! 

We note, however, that the benefits from multi-band data illustrated in Fig.~\ref{fig:PE} do decrease as the total mass  decreases. 
For stellar-mass sources like GW150914, the SNR in the ground-based detectors is so high~\citep{Vitale:2018nif} that little is gained by folding-in LISA's measurements, at least where the binary's parameters are concerned. However, even for stellar mass binaries, combining the low- and high-frequency signals can significantly strengthen some tests of general relativity, as discussed below.

\noindent
{\bf 4) Multiband observations are expected to yield stringent tests of general relativity.} This is for basically the same reason that multiband observations can lead to large improvements in parameter estimation: tracking the GW phase over several decades in frequency 
provides a long "lever-arm" for comparing measurement to theory. The improvements in bounds on various theories of gravity coming from multiband observations have been quantified~\citep{2016PhRvL.116x1104B,2017PhRvD..96h4039C,Berti:2018cxi}; to summarize, IMBH binaries with total mass  $M_{\rm total} \sim 10^3 M_{\odot}$ should provide especially strong constraints, since in that mass regime the SNR can be large in both frequency bands.
The optical configurations of some ground-based instruments could also be optimized, specifically targeting specific features of the signal. For example, Ref.~\cite{Tso:2018pdv} showed that narrow-band setups for current and future detectors might improve ringdown tests of the Kerr nature of BHs by up to a factor of two. The gain from narrow-banding may be larger for third-generation detectors, where quantum noise is expected to dominate other noise sources by a factor $\gtrsim 2$ for
frequencies $\gtrsim 40$~Hz.

\section{Concluding Thoughts} \label{sec:conclusion}
We have sketched just a few examples of how multiband GW data will advance our understanding of stellar-mass BH binaries and IMBHs, beyond what is achievable from either space-based or ground-based GW data alone.  
While we have focused on the science from binaries that are observed in both the high- and low-frequency bands, we would be remiss not to at least mention the science yield from comparing the {\it entire population}  of stellar-mass and intermediate-mass BH binaries observed by LISA with the {\it entire population} of these, as observed by ground-based detectors -- including the sources not observed in both bands within a few years of each other.  
The distribution of binaries in GW frequency reflects the physical processes underlying their creation and evolution (e.g, whether tight binaries are formed dynamically through three-body interactions, or through two-body stellar evolution, would lead to different initial eccentricities, and hence different frequency distributions). 
Observations of the population distribution across the spectrum should provide strong constraints on BH binary formation channels~\citep{2016ApJ...830L..18B,Nishizawa:2016eza,Rodriguez:2017pec,Randall:2018nud,DOrazio:2018jnv,Samsing:2018isx}.
Additionally, this white paper has focused on the enhanced science made available by combining LISA data with ground-based data, but we would be remiss not to mention that there is significant interest internationally in flying a GW mission that would resemble a shorter-armed LISA, with optimal sensitivity in the $\sim 10^{-1}-1\,$Hz range. Currently this interest is strongest in Japan, where the most relevant mission is called B-Decigo~\cite{Sato_2017}, and in China. 
Clearly the synergies explored here would only be enhanced by the ability to follow a single binary's GW emission across the entirety of the range $\sim 10^{-3} - 100\,$Hz.  An extra advantage of adding B-Decigo to our mix of observatories (emphasized in~\cite{2018PTEP.2018g3E01I}) is that B-Decigo would give advance warnings to ground-based detectors of neutron-star/neutron-star (NS/NS) mergers. LISA cannot provide these, since the NS/NS inspiral time, from the LISA band to merger, is far too long.

The theoretical study of multiband GW astronomy is a new and exciting area of research, and we anticipate new ideas %
as such synergies are explored further. 

\newpage
\bibliographystyle{aasjournal}
\bibliography{master}

\begin{thebibliography}{}
\expandafter\ifx\csname natexlab\endcsname\relax\def\natexlab#1{#1}\fi
\providecommand{\url}[1]{\href{#1}{#1}}
\providecommand{\dodoi}[1]{doi:~\href{http://doi.org/#1}{\nolinkurl{#1}}}
\providecommand{\doeprint}[1]{\href{http://ascl.net/#1}{\nolinkurl{http://ascl.net/#1}}}
\providecommand{\doarXiv}[1]{\href{https://arxiv.org/abs/#1}{\nolinkurl{https://arxiv.org/abs/#1}}}

\bibitem[{Abbott {et~al.}(2017)}]{Evans:2016mbw}
Abbott, B.~P., {et~al.} 2017, Class. Quant. Grav., 34, 044001,
  \dodoi{10.1088/1361-6382/aa51f4}

\bibitem[{Apostolatos {et~al.}(1994)Apostolatos, Cutler, Sussman, \&
  Thorne}]{PhysRevD.49.6274}
Apostolatos, T.~A., Cutler, C., Sussman, G.~J., \& Thorne, K.~S. 1994, Phys.
  Rev. D, 49, 6274, \dodoi{10.1103/PhysRevD.49.6274}

\bibitem[{{Baird} {et~al.}(2013){Baird}, {Fairhurst}, {Hannam}, \&
  {Murphy}}]{2013PhRvD..87b4035B}
{Baird}, E., {Fairhurst}, S., {Hannam}, M., \& {Murphy}, P. 2013, \prd, 87,
  024035, \dodoi{10.1103/PhysRevD.87.024035}

\bibitem[{Baker {et~al.}(2019)Baker, Haiman, Rossi, Berger, Brandt, Breedt,
  Breivik, Charisi, Ford, Greene, Hill, Kocsis, Kupfer, Madau, Marsh, McKernan,
  Nissake, Noble, Phinney, Ramsay, Schnittman, Sesana, Stone, Toonen,
  Trakhtenbrot, Vikhlinin, \& Volonteri}]{BakerWP}
Baker, J., Haiman, Z., Rossi, E.~M., {et~al.} 2019, submission to the 2020-2030
  Astronomy and Astrophysics Decadal Survey (Astro2020)

\bibitem[{{Barausse} {et~al.}(2016){Barausse}, {Yunes}, \&
  {Chamberlain}}]{2016PhRvL.116x1104B}
{Barausse}, E., {Yunes}, N., \& {Chamberlain}, K. 2016, Physical Review
  Letters, 116, 241104, \dodoi{10.1103/PhysRevLett.116.241104}

\bibitem[{Bellovary {et~al.}(2019)Bellovary, Colpi, Eracleous, Hornschemeier,
  Mayer, Natarajan, Slutsky, \& Tremmel}]{BellovaryWP}
Bellovary, J., Colpi, M., Eracleous, M., {et~al.} 2019, submission to the
  2020-2030 Astronomy and Astrophysics Decadal Survey (Astro2020)

\bibitem[{Berry {et~al.}(2019)Berry, Hughes, Sopuerta, Chua, Heffernan, Miller,
  \& Sesana}]{BerryWP}
Berry, C. P.~L., Hughes, S.~A., Sopuerta, C.~F., {et~al.} 2019, submission to
  the 2020-2030 Astronomy and Astrophysics Decadal Survey (Astro2020)

\bibitem[{Berti {et~al.}(2018)Berti, Yagi, \& Yunes}]{Berti:2018cxi}
Berti, E., Yagi, K., \& Yunes, N. 2018, Gen. Rel. Grav., 50, 46,
  \dodoi{10.1007/s10714-018-2362-8}

\bibitem[{Berti {et~al.}(2019)Berti, Shoemaker, Barausse, Cholis,
  Garc\'ia-Bellidoc, Hughes, Kelly, Kovetz, Littenberg, Livas, Schnittman, \&
  Yunes}]{BertiWP}
Berti, E., Shoemaker, D., Barausse, E., {et~al.} 2019, submission to the
  2020-2030 Astronomy and Astrophysics Decadal Survey (Astro2020)

\bibitem[{{Breivik} {et~al.}(2016){Breivik}, {Rodriguez}, {Larson}, {Kalogera},
  \& {Rasio}}]{2016ApJ...830L..18B}
{Breivik}, K., {Rodriguez}, C.~L., {Larson}, S.~L., {Kalogera}, V., \& {Rasio},
  F.~A. 2016, \apjl, 830, L18, \dodoi{10.3847/2041-8205/830/1/L18}

\bibitem[{Caldwell {et~al.}(2019)Caldwell, Amin, Hogaa, Holz, Jetzer, Kovetz,
  Smith, \& Tamanini}]{CaldwellWP}
Caldwell, R., Amin, M., Hogaa, C., {et~al.} 2019, submission to the 2020-2030
  Astronomy and Astrophysics Decadal Survey (Astro2020)

\bibitem[{{Chamberlain} \& {Yunes}(2017)}]{2017PhRvD..96h4039C}
{Chamberlain}, K., \& {Yunes}, N. 2017, \prd, 96, 084039,
  \dodoi{10.1103/PhysRevD.96.084039}

\bibitem[{Cholis {et~al.}(2016)Cholis, Kovetz, Ali-Haïmoud, Bird,
  Kamionkowski, Muñoz, \& Raccanelli}]{Cholis:2016kqi}
Cholis, I., Kovetz, E.~D., Ali-Haïmoud, Y., {et~al.} 2016, Phys. Rev., D94,
  084013, \dodoi{10.1103/PhysRevD.94.084013}

\bibitem[{Colpi {et~al.}(2019)Colpi, Holley-Bockelmann, Bogdanovi\'c,
  Natarajan, Sesana, Tremmel, Comerford, Barausse, Berti, Volonteri, Khan, \&
  McWilliams}]{ColpiWP}
Colpi, M., Holley-Bockelmann, K., Bogdanovi\'c, T., {et~al.} 2019, submission
  to the 2020-2030 Astronomy and Astrophysics Decadal Survey (Astro2020)

\bibitem[{Connaughton {et~al.}(2016)}]{Connaughton:2016umz}
Connaughton, V., {et~al.} 2016, Astrophys. J., 826, L6,
  \dodoi{10.3847/2041-8205/826/1/L6}

\bibitem[{Cornish(2019)}]{CornishWP}
Cornish, N. 2019, submission to the 2020-2030 Astronomy and Astrophysics
  Decadal Survey (Astro2020)

\bibitem[{D'Orazio \& Samsing(2018)}]{DOrazio:2018jnv}
D'Orazio, D.~J., \& Samsing, J. 2018, \dodoi{10.1093/mnras/sty2568}

\bibitem[{Eracleous {et~al.}(2019)Eracleous, Gezari, Sesana, Bogdanovic,
  MacLeod, Roth, \& Da}]{EracleousWP}
Eracleous, M., Gezari, S., Sesana, A., {et~al.} 2019, submission to the
  2020-2030 Astronomy and Astrophysics Decadal Survey (Astro2020)

\bibitem[{Farr {et~al.}(2018)Farr, Holz, \& Farr}]{Farr:2017gtv}
Farr, B., Holz, D.~E., \& Farr, W.~M. 2018, Astrophys. J., 854, L9,
  \dodoi{10.3847/2041-8213/aaaa64}

\bibitem[{Farr {et~al.}(2017)Farr, Stevenson, Coleman~Miller, Mandel, Farr, \&
  Vecchio}]{Farr:2017uvj}
Farr, W.~M., Stevenson, S., Coleman~Miller, M., {et~al.} 2017, Nature, 548,
  426, \dodoi{10.1038/nature23453}

\bibitem[{{Farrell} {et~al.}(2009){Farrell}, {Webb}, {Barret}, {Godet}, \&
  {Rodrigues}}]{2009Natur.460...73F}
{Farrell}, S.~A., {Webb}, N.~A., {Barret}, D., {Godet}, O., \& {Rodrigues},
  J.~M. 2009, \nat, 460, 73, \dodoi{10.1038/nature08083}

\bibitem[{Gerosa \& Berti(2017)}]{Gerosa:2017kvu}
Gerosa, D., \& Berti, E. 2017, Phys. Rev., D95, 124046,
  \dodoi{10.1103/PhysRevD.95.124046}

\bibitem[{Gerosa {et~al.}(2018)Gerosa, Berti, O'Shaughnessy, Belczynski,
  Kesden, Wysocki, \& Gladysz}]{Gerosa:2018wbw}
Gerosa, D., Berti, E., O'Shaughnessy, R., {et~al.} 2018, Phys. Rev., D98,
  084036, \dodoi{10.1103/PhysRevD.98.084036}

\bibitem[{Gerosa {et~al.}(2013)Gerosa, Kesden, Berti, O'Shaughnessy, \&
  Sperhake}]{Gerosa:2013laa}
Gerosa, D., Kesden, M., Berti, E., O'Shaughnessy, R., \& Sperhake, U. 2013,
  Phys. Rev., D87, 104028, \dodoi{10.1103/PhysRevD.87.104028}

\bibitem[{{Gerosa} {et~al.}(2019){Gerosa}, {Ma}, {Wong}, {Berti},
  {O'Shaughnessy}, {Chen}, \& {Belczynski}}]{2019arXiv190200021G}
{Gerosa}, D., {Ma}, S., {Wong}, K. W.~K., {et~al.} 2019, arXiv e-prints,
  arXiv:1902.00021.
\newblock \doarXiv{1902.00021}

\bibitem[{Hannam {et~al.}(2014)Hannam, Schmidt, Boh{\'e}, Haegel, Husa, Ohme,
  Pratten, \& P{\"u}rrer}]{Hannam:2013oca}
Hannam, M., Schmidt, P., Boh{\'e}, A., {et~al.} 2014, Phys. Rev. Lett., 113,
  151101, \dodoi{10.1103/PhysRevLett.113.151101}

\bibitem[{{Isoyama} {et~al.}(2018){Isoyama}, {Nakano}, \&
  {Nakamura}}]{2018PTEP.2018g3E01I}
{Isoyama}, S., {Nakano}, H., \& {Nakamura}, T. 2018, Progress of Theoretical
  and Experimental Physics, 2018, 073E01, \dodoi{10.1093/ptep/pty078}

\bibitem[{Littenberg {et~al.}(2019)Littenberg, Breivik, Brown, Eracleous,
  Hermes, Kremer, Kupfer, \& Larson}]{LittenbergWP}
Littenberg, T.~B., Breivik, K., Brown, W.~R., {et~al.} 2019, submission to the
  2020-2030 Astronomy and Astrophysics Decadal Survey (Astro2020)

\bibitem[{McWilliams(2019)}]{McWilliamsWP}
McWilliams, S.~T. 2019, submission to the 2020-2030 Astronomy and Astrophysics
  Decadal Survey (Astro2020)

\bibitem[{Natarajan {et~al.}(2019)Natarajan, Ricarte, Baldassare, Bellovary,
  Bender, Berti, Cappelluti, Ferrara, Greene, Haiman, Holley-Bockelmann,
  Pacucci, Shoemaker, Tremmel, Urry, Vikhlinin, \& Volonteri}]{NatarajanWP}
Natarajan, P., Ricarte, A., Baldassare, V., {et~al.} 2019, submission to the
  2020-2030 Astronomy and Astrophysics Decadal Survey (Astro2020)

\bibitem[{Nishizawa {et~al.}(2016)Nishizawa, Berti, Klein, \&
  Sesana}]{Nishizawa:2016jji}
Nishizawa, A., Berti, E., Klein, A., \& Sesana, A. 2016, Phys. Rev., D94,
  064020, \dodoi{10.1103/PhysRevD.94.064020}

\bibitem[{Nishizawa {et~al.}(2017)Nishizawa, Sesana, Berti, \&
  Klein}]{Nishizawa:2016eza}
Nishizawa, A., Sesana, A., Berti, E., \& Klein, A. 2017, Mon. Not. Roy. Astron.
  Soc., 465, 4375, \dodoi{10.1093/mnras/stw2993}

\bibitem[{{Pasham} {et~al.}(2014){Pasham}, {Strohmayer}, \&
  {Mushotzky}}]{2014Natur.513...74P}
{Pasham}, D.~R., {Strohmayer}, T.~E., \& {Mushotzky}, R.~F. 2014, \nat, 513,
  74, \dodoi{10.1038/nature13710}

\bibitem[{Peters(1964)}]{Peters:1964zz}
Peters, P.~C. 1964, Phys. Rev., 136, B1224, \dodoi{10.1103/PhysRev.136.B1224}

\bibitem[{Punturo {et~al.}(2010)}]{Punturo:2010zz}
Punturo, M., {et~al.} 2010, Class. Quant. Grav., 27, 194002,
  \dodoi{10.1088/0264-9381/27/19/194002}

\bibitem[{P\"urrer {et~al.}(2016)P\"urrer, Hannam, \&
  Ohme}]{PhysRevD.93.084042}
P\"urrer, M., Hannam, M., \& Ohme, F. 2016, Phys. Rev. D, 93, 084042,
  \dodoi{10.1103/PhysRevD.93.084042}

\bibitem[{Randall \& Xianyu(2018)}]{Randall:2018nud}
Randall, L., \& Xianyu, Z.-Z. 2018, Astrophys. J., 864, 134,
  \dodoi{10.3847/1538-4357/aad7fe}

\bibitem[{Robson {et~al.}(2018)Robson, Cornish, \& Liu}]{Cornish:2018dyw}
Robson, T., Cornish, N., \& Liu, C. 2018.
\newblock \doarXiv{1803.01944}

\bibitem[{Rodriguez {et~al.}(2018)Rodriguez, Amaro-Seoane, Chatterjee, \&
  Rasio}]{Rodriguez:2017pec}
Rodriguez, C.~L., Amaro-Seoane, P., Chatterjee, S., \& Rasio, F.~A. 2018, Phys.
  Rev. Lett., 120, 151101, \dodoi{10.1103/PhysRevLett.120.151101}

\bibitem[{Rodriguez {et~al.}(2016)Rodriguez, Zevin, Pankow, Kalogera, \&
  Rasio}]{Rodriguez:2016vmx}
Rodriguez, C.~L., Zevin, M., Pankow, C., Kalogera, V., \& Rasio, F.~A. 2016,
  Astrophys. J., 832, L2, \dodoi{10.3847/2041-8205/832/1/L2}

\bibitem[{Samsing \& D'Orazio(2018)}]{Samsing:2018isx}
Samsing, J., \& D'Orazio, D.~J. 2018, \dodoi{10.1093/mnras/sty2334}

\bibitem[{Sato {et~al.}(2017)Sato, Kawamura, Ando, Nakamura, Tsubono, Araya,
  Funaki, Ioka, Kanda, Moriwaki, Musha, Nakazawa, Numata, ichiro Sakai, Seto,
  Takashima, Tanaka, Agatsuma, suke Aoyanagi, Arai, Asada, Aso, Chiba,
  Ebisuzaki, Ejiri, Enoki, Eriguchi, Fujimoto, Fujita, Fukushima, Futamase,
  Ganzu, Harada, Hashimoto, Hayama, Hikida, Himemoto, Hirabayashi, Hiramatsu,
  Hong, Horisawa, Hosokawa, Ichiki, Ikegami, Inoue, Ishidoshiro, Ishihara,
  Ishikawa, Ishizaki, Ito, Itoh, Kawashima, Kawazoe, Kishimoto, Kiuchi,
  Kobayashi, Kohri, Koizumi, Kojima, Kokeyama, Kokuyama, Kotake, Kozai, Kudoh,
  Kunimori, Kuninaka, Kuroda, ichi Maeda, Matsuhara, Mino, Miyakawa, Miyoki,
  Morimoto, Morioka, Morisawa, Mukohyama, Nagano, Naito, Nakamura, Nakano,
  Nakao, Nakasuka, Nakayama, Nishida, Nishiyama, Nishizawa, Niwa, Noumi,
  Obuchi, Ohashi, Ohishi, Ohkawa, Okada, Onozato, Oohara, Sago, Saijo,
  Sakagami, Sakata, Sasaki, Sato, Shibata, Shinkai, Somiya, Sotani, Sugiyama,
  Suwa, Suzuki, Tagoshi, Takahashi, Takahashi, Takahashi, Takahashi, Takahashi,
  Takahashi, Takahashi, Akiteru, Takano, Taniguchi, Taruya, Tashiro, Torii,
  Toyoshima, Tsujikawa, Tsunesada, Ueda, ichi Ueda, Utashima, Wakabayashi,
  Yamakawa, Yamamoto, Yamazaki, Yokoyama, Yoo, Yoshida, \& Yoshino}]{Sato_2017}
Sato, S., Kawamura, S., Ando, M., {et~al.} 2017, Journal of Physics: Conference
  Series, 840, 012010, \dodoi{10.1088/1742-6596/840/1/012010}

\bibitem[{{Schmidt} {et~al.}(2015){Schmidt}, {Ohme}, \&
  {Hannam}}]{2015PhRvD..91b4043S}
{Schmidt}, P., {Ohme}, F., \& {Hannam}, M. 2015, \prd, 91, 024043,
  \dodoi{10.1103/PhysRevD.91.024043}

\bibitem[{{Sesana}(2016)}]{2016PhRvL.116w1102S}
{Sesana}, A. 2016, Physical Review Letters, 116, 231102,
  \dodoi{10.1103/PhysRevLett.116.231102}

\bibitem[{Stevenson {et~al.}(2017)Stevenson, Berry, \&
  Mandel}]{Stevenson:2017dlk}
Stevenson, S., Berry, C. P.~L., \& Mandel, I. 2017, Mon. Not. Roy. Astron.
  Soc., 471, 2801, \dodoi{10.1093/mnras/stx1764}

\bibitem[{{Taylor} \& {Gerosa}(2018)}]{2018PhRvD..98h3017T}
{Taylor}, S.~R., \& {Gerosa}, D. 2018, \prd, 98, 083017,
  \dodoi{10.1103/PhysRevD.98.083017}

\bibitem[{Tso {et~al.}(2018)Tso, Gerosa, \& Chen}]{Tso:2018pdv}
Tso, R., Gerosa, D., \& Chen, Y. 2018.
\newblock \doarXiv{1807.00075}

\bibitem[{{Veitch} {et~al.}(2015){Veitch}, {Raymond}, {Farr}, {Farr}, {Graff},
  {Vitale}, {Aylott}, {Blackburn}, {Christensen}, {Coughlin}, {Del Pozzo},
  {Feroz}, {Gair}, {Haster}, {Kalogera}, {Littenberg}, {Mandel},
  {O'Shaughnessy}, {Pitkin}, {Rodriguez}, {R{\"o}ver}, {Sidery}, {Smith}, {Van
  Der Sluys}, {Vecchio}, {Vousden}, \& {Wade}}]{2015PhRvD..91d2003V}
{Veitch}, J., {Raymond}, V., {Farr}, B., {et~al.} 2015, \prd, 91, 042003,
  \dodoi{10.1103/PhysRevD.91.042003}

\bibitem[{{Vitale}(2016)}]{2016PhRvL.117e1102V}
{Vitale}, S. 2016, Physical Review Letters, 117, 051102,
  \dodoi{10.1103/PhysRevLett.117.051102}

\bibitem[{Vitale {et~al.}(2017{\natexlab{a}})Vitale, Lynch, Raymond, Sturani,
  Veitch, \& Graff}]{Vitale:2016avz}
Vitale, S., Lynch, R., Raymond, V., {et~al.} 2017{\natexlab{a}}, Phys. Rev. D,
  95, 064053, \dodoi{10.1103/PhysRevD.95.064053}

\bibitem[{Vitale {et~al.}(2017{\natexlab{b}})Vitale, Lynch, Sturani, \&
  Graff}]{Vitale:2015tea}
Vitale, S., Lynch, R., Sturani, R., \& Graff, P. 2017{\natexlab{b}}, Class.
  Quant. Grav., 34, 03LT01, \dodoi{10.1088/1361-6382/aa552e}

\bibitem[{Vitale {et~al.}(2014)Vitale, Lynch, Veitch, Raymond, \&
  Sturani}]{PhysRevLett.112.251101}
Vitale, S., Lynch, R., Veitch, J., Raymond, V., \& Sturani, R. 2014, Phys. Rev.
  Lett., 112, 251101, \dodoi{10.1103/PhysRevLett.112.251101}

\bibitem[{Vitale \& Whittle(2018)}]{Vitale:2018nif}
Vitale, S., \& Whittle, C. 2018, Phys. Rev., D98, 024029,
  \dodoi{10.1103/PhysRevD.98.024029}

\bibitem[{{Wong} {et~al.}(2018){Wong}, {Kovetz}, {Cutler}, \&
  {Berti}}]{2018arXiv180808247W}
{Wong}, K.~W.~K., {Kovetz}, E.~D., {Cutler}, C., \& {Berti}, E. 2018, ArXiv
  e-prints.
\newblock \doarXiv{1808.08247}

\end{thebibliography}
\end{document}